\documentclass[aps,pra,floatfix,twocolumn,10pt,superscriptaddress,showpacs,preprintnumbers,longbibliography]{revtex4-2}
\usepackage[english]{babel}
\usepackage[utf8]{inputenc}
\usepackage{dcolumn}    
\usepackage{bm} 
\usepackage{graphicx}
\usepackage{amsmath}    
\usepackage{enumitem}
\usepackage{latexsym}
\usepackage{amsfonts}   
\usepackage{amssymb}
\usepackage{array}      
\usepackage{epsfig}
\usepackage{braket}
\usepackage{color}
\usepackage[colorlinks=true,linkcolor=blue,urlcolor=blue,citecolor=blue,pdfusetitle]{hyperref}
\usepackage{hyperref}
\usepackage[T1]{fontenc}
\usepackage[sc]{mathpazo}
\newcommand{\ignore}[1]{}

\newcommand{\eq}{Eq.\,}
\newcommand{\eqs}{Eqs.\,}
\newcommand{\fig}{Fig.\,}
\newcommand{\figs}{Figs.\,}
\newcommand{\cf} {cf.~}

\newcommand{\ie} {i.e.~}
\newcommand{\eg} {e.g.~}
\newcommand{\rref} {Ref.\,}
\newcommand{\rrefs} {Refs.\,}

\begin{document}

	\title{Non-Markovian dynamics of a qubit due to accelerated light in a lattice}

	\author{Marcel Augusto Pinto}
	\affiliation{Universit$\grave{a}$  degli Studi di Palermo, Dipartimento di Fisica e Chimica -- Emilio Segr$\grave{e}$, via Archirafi 36, I-90123 Palermo, Italy}
	\author{Giovanni Luca Sferrazza}
	\affiliation{Universit$\grave{a}$  degli Studi di Palermo, Dipartimento di Fisica e Chimica -- Emilio Segr$\grave{e}$, via Archirafi 36, I-90123 Palermo, Italy}
	\author{Daniele De Bernardis}
	\affiliation{	National Institute of Optics (CNR-INO), Via Nello Carrara 1, Sesto Fiorentino, 50019, Italy}
	\affiliation{European Laboratory for Non-Linear Spectroscopy (LENS), Via Nello Carrara 1, Sesto F.no 50019, Italy}
	\author{Francesco Ciccarello}
	\affiliation{Universit$\grave{a}$  degli Studi di Palermo, Dipartimento di Fisica e Chimica -- Emilio Segr$\grave{e}$, via Archirafi 36, I-90123 Palermo, Italy}
	\affiliation{NEST, Istituto Nanoscienze-CNR, Piazza S. Silvestro 12, 56127 Pisa, Italy}
	
	\date{\today}
	
	\begin{abstract}
		We investigate the emission of a qubit weakly coupled to a one-band coupled-cavity array where, due to an engineered gradient in the cavity frequencies, photons are effectively accelerated by a synthetic force $F$. For strong $F$, a reversible emission described by an effective Jaynes-Cummings model occurs, causing a chiral time-periodic excitation of an extensive region of the array, either to the right or to left of the qubit depending on its frequency. For weak values of $F$ instead, a complex non-Markovian decay with revivals shows up. This is reminiscent of dynamics induced by mirrors in standard waveguides, despite the absence of actual mirrors, and can be attributed to the finite width of the energy band which confine the motion of the emitted photon. In a suitable regime, the decay is well described by a delay differential equation formally analogous to the one governing the decay of an atom in a multi-mode cavity where the cavity length and time taken by a photon to travel between the two mirrors are now embodied by the amplitude and period of Bloch oscillations, respectively.
	\end{abstract}

	\maketitle

	\section{Introduction}
	
	The study of qubits coupled to tailored low-dimensional photonic environments, mostly at the few-photon level, lies at the forefront of modern quantum optics, and especially waveguide QED \cite{Chang_RevModPhys.90.031002,Sheremet-RMP,ciccarello2024waveguide}, with potential applications for the effective processing of quantum information. Relying on the progress of experimental capabilities which nowadays enable the fabrication of unconventional photonic baths, \eg lattices or 1D continuous waveguides, coupled to qubits in platforms such as photonic crystals in the optical domain \cite{kimble_pnas.1603788113}, superconducting circuits in the microwaves \cite{Simon_chiralQuantumOptics_2022,gasparinetti_PhysRevX.12.031036, Painter_PhysRevX.11.011015} and matter-wave emulators \cite{Schneble_nature2018_matterwaves, Schneble_PhysRevResearch.2.043307}, the search for novel qubit-photon interaction paradigms is gaining momentum. 
	\begin{figure}[t!]
		\centering
		\includegraphics[width=0.45 \textwidth]{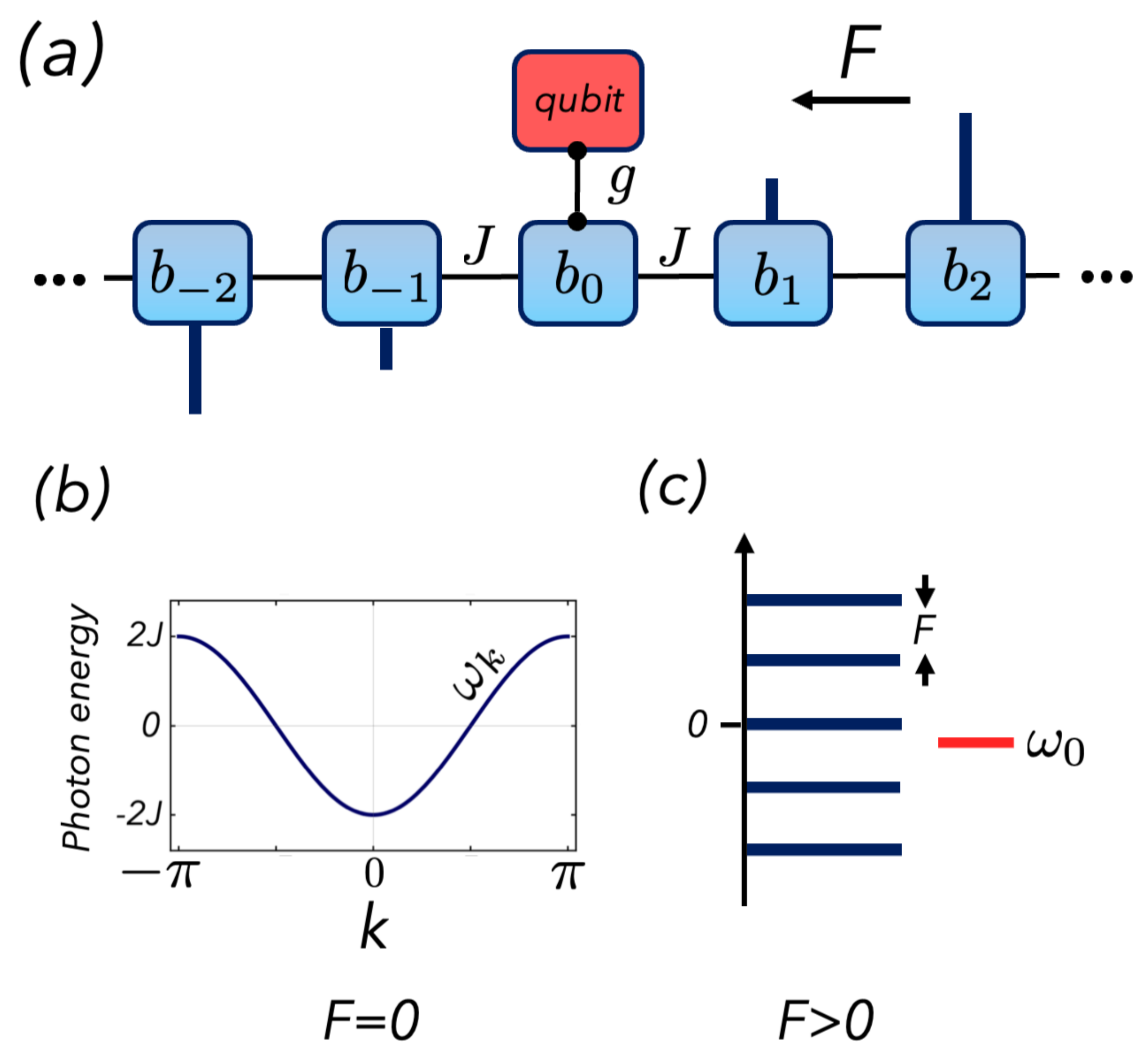}
		\caption{Setup and field's energy spectrum. (a) Qubit emitting into a 1D array of coupled cavities, where $g$ and $J$ are the qubit-cavity coupling strength and photon hopping rate, respectively. Cavity frequencies (vertical bars) are non-uniform thus mimicking a linear potential with corresponding force of strength $F$. (b) Single-photon energy spectrum of the cavity array (bath) for $F=0$, in which case it is a continuous finite band. The quasi-momentum $k$ lies in the first Brillouin zone. (c) Single-photon energy spectrum for $F>0$: the spectrum now is discrete (forming the so called Wannier-Stark ladder) with energy spacing $F$. We also display the qubit frequency $\omega_0$.}
		\label{fig:setup}
	\end{figure}
	
	Within this general framework and concerning emission properties, two major classes of phenomena that do not occur in conventional quantum optics are receiving considerable attention, both at the theoretical and experimental level. One class explores the effect of a {\it band structure} of the photonic bath, which can cause effects such as fractional decay \cite{Sanchez_1993_PhysRevA.47.3380, Tran_1994_PhysRevA.50.1764,Lambropoulos2000a,Schneble_nature2018_matterwaves}, qubit-photon bound states \cite{Calajo_PhysRevA.93.033833, Cirac_PhysRevX.6.021027, Liu2017a} or dipole-dipole dispersive interactions with tunable-range  \cite{Douglas2015b,Tudela2015sub,Sundaresan2019,gasparinetti_PhysRevX.12.031036,zhang2023superconducting}. These phenomena require implementing a {\it periodic} photonic bath, namely a lattice, \eg through an array of coupled cavities or resonators, in order to endow its spectrum with bands and bandgaps. Another important class of  investigated phenomena focus on the non-Markovian regime where photon retardation times (time delays) become comparable or larger than the qubit decay time \cite{DornerPRA02,TufarelliPRA13,GrimsmoPRL15,pichlerPhotonic2016}, which can give rise to qubit revivals \cite{andersson2019non,PainterPRX21}, photon trapping \cite{Calajo2019}, enhanced Dicke superradiance \cite{sinha2020non} or photonic cluster states \cite{PichlerPNAS17}. These phenomena typically require implementing a one-dimensional photonic bath, which could also be just a standard continuous waveguide or transmission line, along with the introduction of a mechanism enforcing the emitted photon to return to the qubit, \eg one or more mirrors, in a time larger than the qubit decay time.
	
	In this work, we investigate for the first time the emission of a qubit into an infinitely long 1D coupled-cavity array where emitted light is accelerated by a synthetic {\it force} [see \fig\ref{fig:setup}(a)], which can be easily implemented in the lab through an engineered gradient of cavity frequencies mimicking a linear potential sensed by the photons. 
    Despite the qubit being weakly coupled to the lattice, based on Bloch oscillations \cite{bloch1929quantenmechanik,zener1934theory} the synthetic force drives the emitted photon along the band until it hits a band edge. This gives rise to a confined motion of emitted light even in real space, despite no actual mirrors are present, resulting in a rich non-Markovian dynamics. We identify in particular two different regimes depending on the ratio between the Bloch oscillations period and the qubit decay time. When this ratio is small (strong-force regime), the system can undergo vacuum Rabi oscillations which are yet chiral in that the qubit periodically exchanges energy with an extensive region of the array which can lie either to its right or left. Notably this chirality can be controlled by tuning the qubit frequency. For large values of the aforementioned ratio (weak-force regime), the qubit instead shows up a non-Markovian decay that features a series of revivals and secondary emitted wavepackets each one undergoing a round trip (before hitting back the qubit) whose duration and traveled distance depend on the force strength.

	The present paper is organized as follows. In Section \ref{sec-model}, we introduce the model and Hamiltonian. In Section \ref{sec-eig-HB}, we review the spectrum of and eigenstates of the bare lattice, where the latter ones have the form of Bessel functions.
    The following Section \ref{sec-BO} is dedicated to a review of the Bloch oscillations.
    In Section \ref{sec-simul}, we present a numerical study of qubit emission in the regimes of strong and weak synthetic force, showing chiral vacuum Rabi oscillations and non-Markovian revivals. 
    In Section \ref{sec-dde}, we develop a theoretical analysis of such phenomena 
    by formulating 
    a specific definition of the two regimes; we explicitly show that a qubit frequency parked at the band center follows a delay differential equation analogous to a multi-mode cavity-QED system. 
    In Section \ref{sec-exp}, we discuss potential experimental tests of the predicted phenomena.
    Finally, in Section \ref{sec-conc}, we draw our conclusions and discuss future directions. 
    Some technical details are presented in the Appendixes.
			\section{Model and Hamiltonian}\label{sec-model}
	
	We consider a two-level qubit (quantum emitter) with ground (excited) state $\ket{g}$ ($\ket{e}$) and frequency $\omega_0$, locally coupled to an infinite 1D array of coupled cavities labeled by integer $n$  [see \fig\ref{fig:setup}(a)]. The total Hamiltonian reads (we set $\hbar=1$ throughout)
	   \begin{equation}
		H=\omega_0 \sigma_+\sigma_-+ H_B+g (b_{n_0}^\dag \sigma_-+{\rm H.c.}) \,,\label{Htot}
	\end{equation}
	where  $H_B$ is the free Hamiltonian of the field (``bath") given by
	  \begin{equation}
		H_B=\sum_n (n F) \,b_n^\dag b_n- J \sum_n  (b_{n+1}^\dag b_n+{\rm H.c.})\,.\label{HB}
	\end{equation}
	Here, $\sigma_-=\sigma_+^\dag=|g\rangle\langle e|$ are usual pseudo-spin ladder operators of the qubit while $b_n$ are bosonic ladder operators of $B$ (one for each cavity). The qubit is locally coupled under the rotating-wave approximation to cavity $n_0$ (``qubit position") with strength $g$ [\cf\eq\eqref{Htot} and \fig\ref{fig:setup}(a) where $n_0=0$]. 
	In \eq\eqref{HB}, the second sum is the standard tight-binding Hamiltonian describing photon hopping between nearest-neighbour cavities with $J>0$ the hopping rate. The first sum instead accounts for the bare Hamiltonian of each single cavity: importantly for the present paper, notice that cavity frequencies are {\it non-uniform} in that they increase linearly with the cavity index $n$ with slope $F$. Accordingly, the first sum in \eq\eqref{HB} is naturally interpreted as an effective scalar potential on the photons whose corresponding {\it force} (opposite of gradient) is measured by $F$. 
	
	The total Hamiltonian conserves the total number of excitations $N_{\rm exc}=\sigma_+\sigma_-+\sum_n b_n^\dag b_n$. In this work, we will be concerned with the single-excitation subspace (i.e., the eigenspace $N_{\rm exc}=1$) this being spanned by $\{\ket{e},\{\ket{n}\}\}$. Here, we conveniently adopted the compact notation such that $\ket{e}\ket{\rm vac}$ (with $\ket{\rm vac}$ the field's vacuum state) is renamed as $\ket{e}$, while each state $\ket{g}b_n^\dag \ket{\rm vac}$ is renamed as $\ket{n}$. Thus $\ket{e}$ is now intended as the joint state of the qubit-field system where one excitation sits on the qubit, while $\ket{n}$ features one excitation on cavity $n$.
	\\
	\\
	We next review the energy spectrum and eigenstates of the bare field Hamiltonian $H_B$, which is an essential step to understand the qubit's decay dynamics.
	
\section{Field spectrum and eigenstates}\label{sec-eig-HB}

Let us first shortly recall that in the special case of zero force (\ie for $F=0$), $H_B$ is translationally invariant and Bloch theorem holds. Accordingly, the field's eigenstates read
\begin{equation}
|k\rangle=\frac{1}{\sqrt{N}} \sum_n e^{i k n} \ket{n}\label{ks}
\end{equation}
with $N\gg 1$ the number of cavities and $k$ the quasi-momentum such that $-\pi<k\le \pi$, this interval being the {\it first Brillouin zone} (FBZ). The photon dispersion law is given by 
\begin{equation}
	\omega_{k}=-2 J \cos k\label{wk}\,.
\end{equation}
For the goals of the present work, it is worth stressing that both the first Brillouin zone (FBZ) and the lattice energy band have {\it finite size}, being respectively 2$\pi$ and $4J$ [see \fig\ref{fig:setup}(b)]. Therefore, although the array is infinite in real space, photons effectively live within a finite-size region in both the quasi-momentum and energy space.

We now consider Hamiltonian \eqref{HB} in the general case where $F$ is non-zero. Unlike conventional waveguide-QED setups, due to the potential term $\propto F$ [\cf\eq\eqref{HB}] in the current system the field does not enjoy translational invariance hence  Bloch theorem does not hold. However, the energy spectrum and stationary states of $H_B$ can still be analytically worked out as (see \eg \rrefs\cite{hacker1970stark,fukuyama1973tightly,Hartmann_2004})
    \begin{align}
    	\omega_n&=n F\,,\label{ladder}\\
    	\ket{\varphi_n}&=\sum_m J_{m-n}(\xi) \ket{m} \,\label{phin}
    \end{align}
    with $H_B \ket{\varphi_n}=\omega_n \ket{\varphi_n}$ for $n$ running over all integers. Here, $J_\alpha(x)$ denotes a Bessel function of the first kind of order $\alpha$ and argument $x$, while $\xi$ is defined as
    \begin{equation}
    \xi=\frac{2J}{F}\label{xi}\,
    \end{equation}
    and from now on will be referred to as the {\it localization length}. It is understood the above spectrum and eigenstates hold for an infinitely long array, which is appropriate here as our goal is to investigate qubit emission in the bulk of the coupled-cavity array.
    
\begin{figure}[t!]
    	\centering
    	\includegraphics[width=0.48\textwidth]{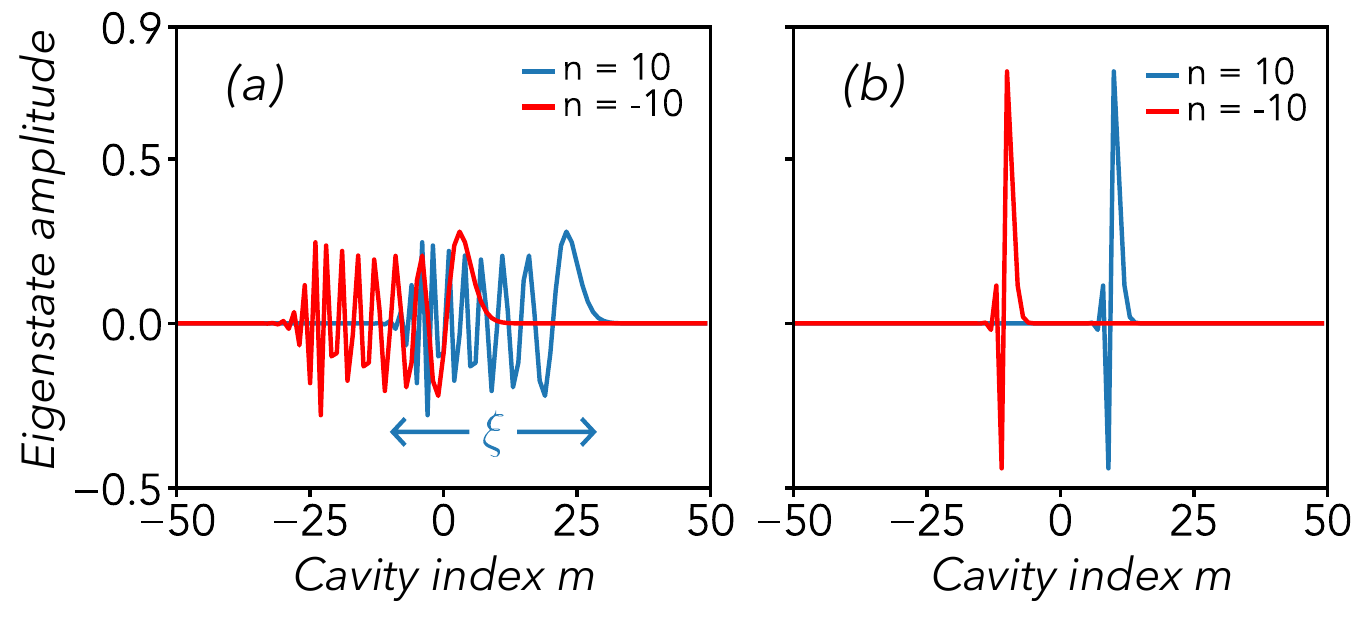}
    	\caption{Wavefunctions of the field's eigenstates $\ket{\varphi_n}$. We report the representative cases $n=-10,10$ for $\xi=15$ (a) and $\xi=1$ (b). Each wavefunction $\langle m\ket{\varphi_n}$ is localized with characteristic localization length $\xi$ around cavity $n$, in the vicinity of which it shows up spatial oscillations before eventually decaying exponentially for $|m-n|\gtrsim \xi$. For $\xi$ large enough, as in panel (a), eigenstates get very delocalized and strongly overlapping.}
    	\label{fig:bessel}
    \end{figure}

    According to \eq\eqref{ladder} and as shown in \fig\ref{fig:setup}(c), the field's spectrum is thus discrete for any $F>0$, consisting of a {\it ladder} structure (known as ``Stark-Wannier ladder") with constant energy spacing given by the force, i.e., $\omega_{n+1}{-}\omega_n {=}F$. Each eigenstate $\ket{\varphi_n}$ has a wavefunction {\it localized} around the $n$th cavity with a characteristic width given by the localization length $\xi$ [\cf \eq\eqref{xi}] (this justifies the choice of label $n$ which intentionally coincides with the cavity index $n$). \fig\ref{fig:bessel} illustrates the spatial profile of states $\ket{\varphi_n}$ in the two representative cases $n=-10,10$ for a large (small) value of $\xi$ in panel (a) [(b)]. Notice that the wavefunction is the same for all states except for an $n$-dependent displacement. For $|m-n|\gtrsim \xi$ the wavefunction exponentially decays with $m$.  Notice that spectrum \eqref{ladder} is just the same that would occur for zero hopping rate, i.e., for $J=0$, in which case the cavities are fully uncoupled from each other with the field's eigenstates trivially reducing to $\ket{\varphi_n}=\ket{n}$ [indeed $\xi=0$ in this case, \cf\eq\eqref{xi}]. \eqs\eqref{ladder} and \eqref{phin} state that as $J>0$ the energy spectrum remains unaffected, however each state is no longer localized on one cavity as a consequence of photon hopping. 
    
    As a property which will be used in Section \ref{sec-dde}, observe that for $\xi$ large enough [see \fig\ref{fig:bessel}(a)] wavefunction $\langle m\ket{\varphi_n}$ displays an oscillatory behaviour in the vicinity of cavity $n$ (i.e, its center). In this central region of its support, the wavefunction of each eigenstate can indeed be approximated by a sinusoid according to
    \begin{equation}
J_{m-n}(\xi) =\sqrt{\frac{2}{\pi \xi}} \,\sin \left[(n-m) \frac{\pi}{2}+\xi+\frac{\pi}{4}\right]\,\,{\rm for} \,\,n\ll \sqrt{\xi}  \label{sin}
    \end{equation}
    which is a standard asymptotic expansion of Bessel functions \cite{abramowitz1965handbook}.

    We also point out that for $\xi\gtrsim 1$ (requiring the force to be weak enough) eigenstates are generally overlapping one another, with the overlap increasing more and more as $\xi$ gets larger. 

     \section{Bloch oscillations}\label{sec-BO}
       \begin{figure*}[t!]
    	\centering
    	\includegraphics[width=0.95 \textwidth]{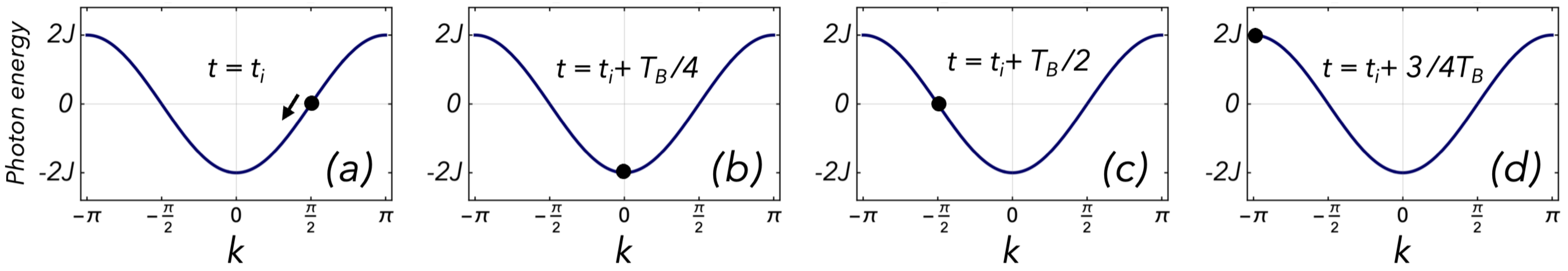}
    	\caption{Skecth of the time evolution of quasi-momentum in the presence of a force (pointing to the left), illustrating Bloch oscillations in the momentum space. We consider the initial condition  $k(t_i)=k_i=\pi/2$ [see panel (a)]. Panels (b), (c) and (d) show the evolved momentum at times $t=t_i+T_B/4$, $t=t_i+T_B/2$ and $t=t_i+3/4T_B$, respectively, where $T_B$ is the period [\cf\eq\eqref{TB}]. At time $t=t_i+T_B$ (not shown), the photon momentum takes again the initial value. Note that the photon velocity $v(t)=d\omega_k/dk$ vanishes whenever $k(t)=0, \pm\pi$ (where the slope of $\omega_{k}$ is zero): at these times the photon hits a band edge and reverses its motion. In the instance considered here, the first motion reversal occurs at time $t=t_i+T_B/4$ when the photon reaches the lower band edge.}
    	\label{fig:BOs}
    \end{figure*}
    In this section, we continue to review bare field properties.
    As said, introducing the force $F$ breaks translational invariance of Hamiltonian $H_B$, hence plane waves \eqref{ks} are no longer stationary states. However, the time evolution of a given plane wave $\ket{k}$ is particularly simple as it can be shown that $e^{-i H_B t}\ket{k}\propto \ket{k(t)}$ with $k(t)$ satisfying $\frac{d k}{d t}= -F$. Hence, if $k=k_i$ at the initial time $t_i$, at a later time 
    \begin{equation}
    k(t)= k_i-F(t-t_i)\,.\label{newton}
    \end{equation}
   
    This evolution is formally analogous to the familiar Newton's second law for a classical particle driven by a constant force undergoing uniformly accelerated motion. In the present {\it lattice}, however, the photon quasi-momentum cannot grow indefinitely since it is constrained to lie in the FBZ $-\pi <k \le \pi$ [\c\fig\ref{fig:setup}(b)]. As a consequence, when $k(t)$ reaches the left boundary of the FBZ, i.e., for $k(t)=-\pi$, it will suddenly reappear on the right FBZ boundary and next resume its linear evolution until reaching again the left one (see skecth in \fig\ref{fig:BOs}). This results in a {\it periodic} evolution, the celebrated ``Bloch oscillations" \cite{bloch1929quantenmechanik,zener1934theory}, whose period based on \eq\eqref{newton} is given by
    \begin{equation}
    T_B=\frac{2\pi}{F}\label{TB}
    \end{equation}
    Notably, unlike a standard massive particle undergoing a spatially unbound uniformly accelerated motion, here the constant force makes the photon perform a {\it bound} motion in real space driven by the time-dependent acceleration $a(t)=(d^2\!\omega_k/dk^2) F $ as a consequence of the periodic shape of the dispersion law [see \eq\eqref{wk} and \fig\ref{fig:setup}(b)].
    Importantly, the photon group velocity $v(t)=d\omega_k/dk$ (slope of the dispersion law) vanishes at the two band edges corresponding to $k(t)=0, \pm\pi$: at these times therefore the photon reverses its motion in real space.

   More in detail, by replacing \eqref{newton} in the group velocity $v_k=d\omega_k/dk=2 J \sin k$ [\cf\eq\eqref{wk}], we get the time-dependent speed of the photon $v(t)=2J \sin [k_i-F (t-t_i)]$. Upon integration and by replacing $F=2\pi/T_B$, this yields that the photon performs a real-space oscillatory motion of period $T_B$ described by
   \begin{equation}
   	x(t)=x_i+\xi \cos k_i-\xi\cos \left(k_i- 2\pi\,\frac{t-t_i}{T_B} \right)\,\label{xt}
   \end{equation}    
   with $x_i=x(t_i)$.
  In particular, we see that the oscillation amplitude coincides with the localization length $\xi$ [\cf\eq\eqref{xi}]. Notice that these real-space oscillations are in phase with the time evolution of the photon energy
  \begin{equation}
  \omega(t)=-2 J \cos \left(k_i- 2\pi\,\frac{t-t_i}{T_B} \right)\,,\label{wt}
  \end{equation}
   which implies that at those times at which the photon is back to the initial position $x_i$ its energy returns to the initial value $\omega(t_i)=-2 J \cos k_i$.

    \section{Dynamics of qubit emission via numerical simulations}\label{sec-simul}

   Our goal in this paper is investigating emission properties of the qubit, i.e., the dynamics arising with the initial state $\ket{\Psi(0)}=\ket{e}$ (qubit excited and field in the vacuum state). The present section illustrates some typical properties of the emission dynamics based on numerical simulations in real and momentum space. 
         \begin{figure*}[t!]
    	\centering
    	\includegraphics[width=0.92 \textwidth]{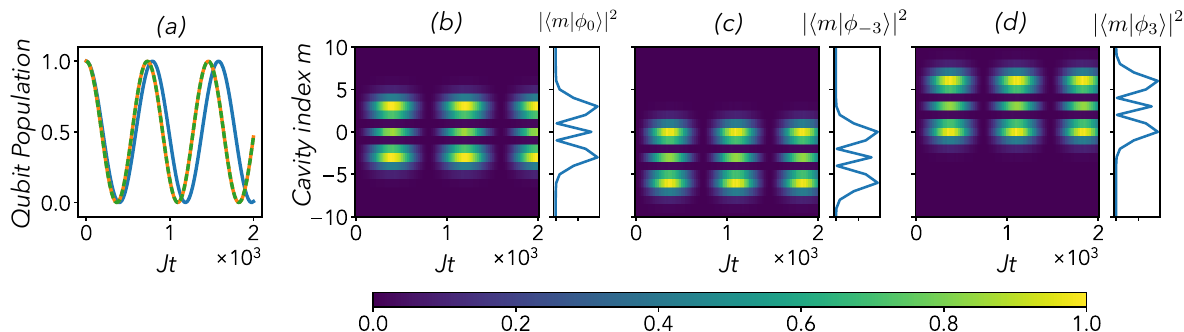}
    	\caption{Vacuum Rabi oscillations under a strong synthetic force. (a) Time behaviour of the qubit excited-state population $|{\alpha_e}|^2$ for $\omega_0=0$ (solid blue curve), $\omega_0=-3 F$ (solid orange) and $\omega_0=3 F$ (dashed). (b) Real-space photon density $|{\beta_n}|^2$ versus time for $\omega_0=0$ [panel (b)], $\omega_0= -3 F$ (c) and $\omega_0= 3 F$ (d). We consider the initial state $\ket{\Psi(t=0)}{=}\ket{e}$ . Throughout we set $n_0=0$, $g=0.01 J$ and $F=0.5 J$. Time is expressed in units of $J^{-1}$. In each of panels (b)-(d), the photon density is normalized to its maximum value for each respective case. To the right of each of panels (b)-(d), we also show for comparison the spatial profile of $|\langle m|\varphi_n\rangle|^2$ for $n=0$ (b), $n=-3$ (c) and $n=3$ (d).}
    	\label{fig:rabi}
    \end{figure*}  
    
    Due to conservation of the total number of excitations, the joint evolved state at any time $t$ can be arranged in the form
    \begin{equation}
    	\ket{\Psi(t)}={\alpha_e}(t) \ket{e}+\sum_n \beta_n(t) \ket{n}\label{psit1}
    \end{equation}
    with ${\alpha_e}(t)$ and $\beta_n(t)$ the probability amplitudes to find the system in state $\ket{e}$ and $\ket{n}$, respectively, at time $t\ge 0$. Representing the propagator $U(t)=e^{-i H t}$, with $H$ given by \eq\eqref{Htot}, in the basis $\{\ket{e},\{\ket{n}\}\}$ gives rise to a square matrix of dimension $N$+1. Once this is applied to the column vector  corresponding to the initial state $\alpha_e(0)=1$ and $\beta_n(0)=0$ for any $n$, one obtains the evolved amplitudes  $\alpha_e(t)$ and $\beta_n(t)$ that fully specify $\ket{\Psi(t)}$. 
    
    An alternative yet useful representation of the dynamics is in terms of the single-photon basis \eqref{ks}. In this picture, the evolved joint state reads
        \begin{equation}
    	\ket{\Psi(t)}={\alpha_e}(t) \ket{e}+\sum_k \gamma_k(t) \ket{k}\label{psit2}\,,
    \end{equation}
    which differs from \eqref{psit1} in that the field is now represented in the momentum space. In Section \ref{sec-NM} we will use both pictures to get insight into the emission dynamics.
    
    Throughout this paper, we will consider values of $g$ [\cf\fig\ref{fig:setup}(a) and \eq\eqref{Htot}] such that $g\ll J$. The rationale of this choice is that for $F=0$ a qubit tuned within the photonic band (having width $\propto J$) is in the standard Markovian regime (a case which we will review at the beginning of Section \ref{sec-NM}). Accordingly, the occurrence of non-Markovian effects, if any, stems from the introduction of the synthetic force which is our main focus in this work. 
    Throughout this section, we will consider the case $n_0=0$, \ie a qubit coupled to cavity $n_0=0$ as in \fig\ref{fig:setup}(a).
    
    In what follows we analyze first chiral vacuum Rabi oscillations (Section \ref{sec-rabi}) and then non-Markovian decay with revivals (Section \ref{sec-NM}), corresponding respectively to the regimes of strong and weak values of force $F$. A rigorous definition of these two regimes will be discussed in Section \ref{sec-regimes}.

    \subsection{Chiral vacuum Rabi oscillations}\label{sec-rabi}
    
As said, for $F>0$ the bare field spectrum is the discrete Wannier-Stark ladder [see \eq\eqref{ladder} and \fig\ref{fig:setup}(c)]. Accordingly, by tuning the qubit close to one of the field normal frequencies $\omega_n{=}n F$ and provided that the force (modes energy spacing) is large enough compared to the effective qubit-mode coupling strength (more on this in Section \ref{sec-regimes}), one expects vacuum Rabi oscillations to occur since the qubit in fact interacts resonantly with only one field eigenstate $\ket{\varphi_n}$. This is indeed the case as shown in \fig\ref{fig:rabi}, where we consider the initial state $\ket{\Psi(0)}{=}\ket{e}$ and plot the time evolution of the qubit excited-state population $|{\alpha_e}(t)|^2$ [panel (a)] and real-space photon density $|{\beta_n}(t)|^2$ [panels (b)-(d)] for $g=0.01 J$, $F=0.5 J$ and the three representative qubit frequencies $\omega_0=0,\pm 3 F$. While the qubit undergoes undamped oscillations, a region of lattice gets periodically populated, which witnesses the expected qubit-field energy swap typical of vacuum Rabi oscillations. Notice that, despite the qubit being directly coupled to cavity $n_0=0$, several cavities in a range of size $\sim \xi$ [\cf \eq\eqref{xi}] get populated according to a pattern that features secondary maxima and minima. Even more interestingly, depending on the qubit frequency, such a region can be significantly displaced from cavity $n_0=0$ (qubit location) either to the left, as in panel (c), or to the right, as in panel (d). This shows the occurrence of chiral reversible emission into the lattice, whose chirality can be controlled by simply tuning the qubit frequency. The structured pattern of the emitted photon density reflects the shape of a specific field eigenstate [\cf\eq\eqref{phin} and \fig\ref{fig:bessel}], specifically $|\varphi_0\rangle$ in the case of panel (b), $|\varphi_{-3}\rangle$ in panel (c) and $|\varphi_{3}\rangle$ in (d). This is confirmed by the behaviour of $|\langle m|\varphi_n\rangle|^2$ for $n=0,\pm 3$ which is displayed for comparison on the right of each of panels (b)-(d) \footnote{Note that this is an even function of $m$ despite $\langle m|\varphi_0\rangle$ is not, which follows from the parity property of Bessel functions $J_{-n}(\xi)=(-1)^n J_n(\xi)$ \cite{abramowitz1965handbook}. For the specific value of $F$ set in this example, \eq\eqref{xi} yields $\xi=4$, which matches the range of the array being populated during the evolution [\cf\fig\ref{fig:rabi}(b)-(d]}.

Interestingly, unlike a number of recent works \cite{bello_doi:10.1126/sciadv.aaw0297, DeBernardis_PhysRevLett.126.103603,de2023chiral,Leonforte_PhysRevLett.126.063601}, the chiral nature of the above dynamics here is not due to topological properties, but rather to the fact that photonic eigenmodes of different frequencies are localized around different cavities of the array.
A more detailed description of vacuum Rabi oscillations will be provided in Section \ref{sec-rabi-th}.

     \subsection{Non-Markovian decay with revivals}\label{sec-NM}
    
    We next investigate situations in which the qubit significantly interacts with a very large number of field eigenstates, which in particular requires $F$ to be weak enough (compared to the coupling strength) so as to make the ladder spectrum dense [recall \eq\eqref{ladder} and \fig\ref{fig:setup}(c)]. Unlike vacuum Rabi oscillations in Section \ref{sec-rabi}, the qubit is now expected to undergo an irreversible decay, although generally non-Markovian.
         \begin{figure*}[t!]
    	\centering
    	\includegraphics[width=0.9 \textwidth]{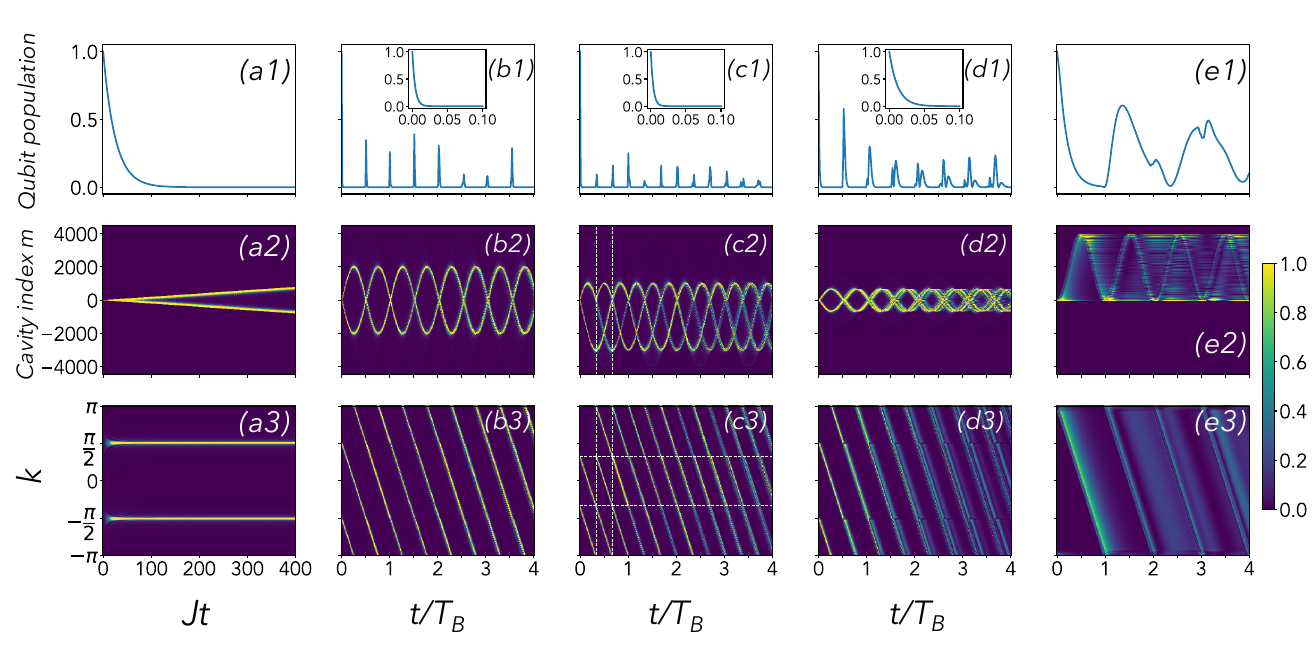}
    	\caption{Emission dynamics under a weak synthetic force. Upper panels show the time behavior of the qubit population $|{\alpha_e}|^2$, while middle (lower) panels show the evolution of photon density in real (quasi-momentum) space $|\beta_m|^2$ ($|\gamma_k|^2$). We set $\omega_0=0$, $g=0.2 J$, $F=0$ [panels (a1)-(a3)], $\omega_0=0$, $g=0.2 J$, $F=10^{-3}J$ [panels (b1)-(b3)], $\omega_0=-J$, $g=0.2 J$, $F=10^{-3}J$ [panels (c1)-(c3)], $\omega_0=0$, $g=0.2 J$, $F=3\cdot10^{-3}J$ [panels (d1)-(d3)] and $\omega_0=1.966 J$, $g = 0.01J$, $F=10^{-3}J$ [panels (e1)-(e3)]. The insets in panels (b1), (c1) and (d1) show a zoom of the qubit decay in the early stages of the dynamics. The vertical dashed lines in (c2) and (c3) mark the return times of the first two emitted wavepackets, while the pair of horizontal dashed lines in (c3) correspond to $k=\pm k_0$. In central and lower panels, photon density is rescaled to its maxixum value. Time is units of $J^{-1}$ in (a1), (a2) and (a3) and of $T_B=2\pi/F$ in all the remaining cases.
    	\label{fig:sim}} 
    \end{figure*}
    In \fig\ref{fig:sim}, we plot the time evolution of the qubit excitation (upper panels) and photon density in real and momentum space (central and lower panels, respectively) for different yet weak values of $F$.
    
    In the trivial case $F=0$ [\cf \figs\ref{fig:sim}(a1)-(a3)], as expected (see \eg \rref \cite{lombardo2014photon}) the qubit undergoes standard Markovian emission with decay rate (for $\omega_0=0$)
    \begin{equation}
    	\Gamma=\frac{g^2}{J}\,.\label{Gamma}
    \end{equation}
    The emitted field [\cf \fig\ref{fig:sim}(a2)] splits into a pair of wavepackets propagating away from the qubit at constant speed, one to the right and one to the left. This is because the qubit couples predominantly to almost resonant modes with $k \simeq \pm k_0$ such that $\omega_{\pm k_0}=\omega_0$, as confirmed by the emitted field evolution in momentum space shown in \fig\ref{fig:sim}(a3). These modes have speed $v_{\pm k_0}=\pm 2J \sin k_0$, which for the considered case $\omega_0=0$ reduce to $v_{\pm k_0}=\pm 2J$ corresponding to the velocity of the right-propagating and left-propagating wavepackets.

We next discuss \figs\ref{fig:sim}(b1)-(c3), where we set $F=10^{-3}J$ while $\omega_0=0$ [panels (b1), (b2), (b3)] and $\omega_0=-J$ [panels (c1), (c2), (c3)]. Clearly, introducing the weak force $F$ dramatically changes the emission dynamics. The qubit population [\cf\figs\ref{fig:sim}(b1) and (c1)] first decays exponentially with rate \eqref{Gamma} and then generally shows up a sequence of partial revivals with the field comprising a series of secondary wavepackets being emitted from the qubit at different times.
 Unlike the case $F=0$, the emitted photon now undergoes an accelerated motion, as witnessed by the non-linear time dependence of the average position of each wavepacket emitted to the left or right during the dynamics [\cf\figs\ref{fig:sim}(b2) and (c2)]: this typically propagates away and slows down until an inversion of motion occurs, and next returns to the qubit at increasing speed. As the wavepacket hits back the qubit, this generally undergoes a partial revival [\cf\figs\ref{fig:sim}(b1) and (c1)]. Interestingly, in striking contrast to the trivial dynamics for $F=0$ [see \fig\ref{fig:sim}(a3)], in momentum space [\cf\figs\ref{fig:sim}(b3) and (c3)] each emitted wavepacket evolves linearly in time in agreement with \eq\eqref{newton} and Section \ref{sec-BO}. This motion is yet constrained owing to the finite size of the FBZ, which enforces a wavepacket reaching the left edge of the FBZ at $k{=}{-}\pi$ to instantaneously reappear from the right edge at $k{=}\pi$. In this momentum space, as an emitted wavepacket of momentum $\pm k_0$ is back to the qubit (return time) its  momentum is reversed (i.e., $\pm k_0\rightarrow \mp k_0$). Notice that, at this return time, a new pair of wavepackets of quasi-momenta $\pm k_0$ are generated, which is more apparent in panel (c3) as highlighted by the horizontal dashed lines showing that the momentum of an emitted wavecpacket is always $k_0$ or $-k_0$ [the dashed vertical lines show the return times of the two wavepackets emitted at time $t=0$ and are also displayed in panel (c2)].
     \begin{figure*}[t!]
		\centering

        \includegraphics[width=0.9 \textwidth]{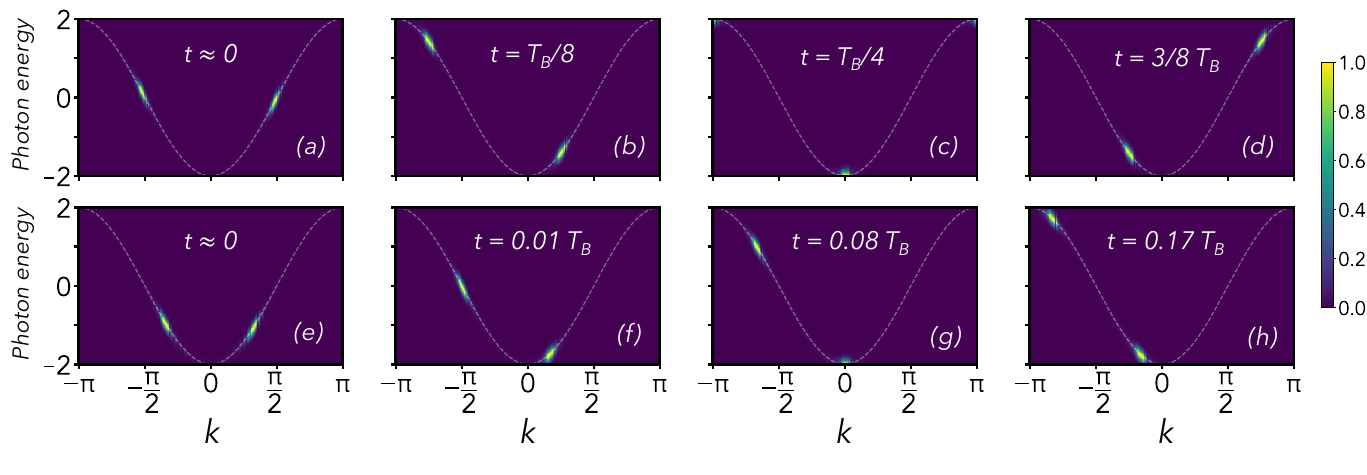}
		\caption{Time behaviour of the photon density on the energy-momentum space for $\omega_0=0$, $g=0.2 J$ and $F=10^{-3} J$ (upper panels) and $\omega_0=-J$, $g=0.2 J$ and $F=10^{-3} J$ (lower panels). At a given time $t$, we numerically compute $|\beta_k(t)|^2$ for each value of $k$ within the FBZ and build up the string $\{k, \omega=-2J \cos k, |\beta_k(t)|^2\}$, which is then used to produce a density plot on the plane $k$-$\omega$. The dashed line shows the photon dispersion law [\cf\fig\ref{fig:setup}(a) and \eq\eqref{wk}], which is superimposed on the density plot for comparison. The photon density is rescaled to its maximum value in each panel.}
		\label{fig:disp}
	\end{figure*}  
	 
The set of {\it return times} $t_r$ at which emitted wavepackets are back to the qubit [see \fig\ref{fig:sim}(b2)-(b3)] result from the combined effect of Bloch oscillations and the fact the qubit mostly couples to almost resonant plane waves $\ket{k(t)}$ such that $\omega_{k(t)}\simeq \omega_0$ (\cf Section \ref{sec-BO}). For $\omega_0{=}0$ [\cf\figs\ref{fig:sim}(b2) and (b3)], $t_r$ are multiple integers of half the Bloch oscillations period, i.e., $t_r=\nu T_B/2$ with $\nu=1,2,...$ [\cf\eq\eqref{TB}]. To see this note that, based on the $k$-space picture in Section \ref{sec-BO} and \fig\ref{fig:BOs}, at $t=0$ the almost resonant momenta are $\pm k_0$ with $k_0= \pi/2$. After a time $T_B/2$, we get $\pm k_0\rightarrow k(t{=}T_B/2)=\mp k_0$: at this time, based on \eqs\eqref{xt} and \eqref{wt} for $x_i=0$ and $k_i=\pm k_0=\pm \pi/2$, either wavepacket is thus back to the initial average position and its energy is again resonant with the qubit which can thus get (partially) re-excited. The qubit next decays back by generating two new wavepackets with $k{\simeq }\pm k_0$ which again undergo an analogous evolution before returning back to the qubit at time $t_r=T_B$ and so on and so forth. 
This is confirmed by \fig\ref{fig:disp}(a)-(d) where, for the same case as in \figs\ref{fig:sim}(b1), (b2) and (b3), we plot the time evolution of the photon density on the energy-momentum space. 
As expected, the emitted wavepackets propagate along the photon dispersion law (dashed curve), which makes them off-resonant with the qubit during the evolution until they both return to the band center thus re-establishing resonance.
This representation additionally makes apparent the confining effect of band edges at $\omega=\pm 2J$ (where  inversions of motion occur), which work as a pair of effective  ``mirrors" but in the energy domain, this being the ultimate reason of the non-Markovian nature of the dynamics.

For $\omega_0$ not lying at the band center (but still far from the band edges), as in \figs\ref{fig:sim}(c1), (c2) and (c3), wavepackets emitted to the right and those to the left now evolve {\it differently} in terms of both the distance traveled up to an inversion of motion and return times, as can be seen from panels (c2) and (c3). This is best understood in the energy-momentum space, displayed in the lower panels of \fig\ref{fig:disp}: since $\omega_0$ is no longer at the band center, the time taken by the emitted wavepacket of initial average momentum $+k_0$ to evolve into $-k_0$ is now shorter than the time taken by the wavepacket $-k_0$ to turn into $+k_0$ (for $0{<}\omega_0{<}2J$ the converse statement would hold) with the two times summing to $T_B/2$. By using again \eqs\eqref{xt} and \eqref{wt}, this entails that the former wavepacket travels a shorter distance before undergoing an inversion of motion and returns earlier to the qubit. 
This is reflected, in particular, by the first two qubit revivals in \fig\ref{fig:sim}(c1), one shorter one longer than $T_B/2$. Since, as said, as a wavepacket is reabsorbed a new pair of wavepackets of opposite momenta $\pm k_0$ are generated, a series of return times of growing density arises as is apparent in \fig\ref{fig:sim}(c2) and (c3), corresponding to an increasingly complex pattern of qubit revivals [see \fig\ref{fig:sim}(c1)]. The series of return times can be predicted as a function of $\omega_0$ by using \eq\eqref{xt}, based on which a wavepacket $\pm k_0$ emitted at time $t_i$ is back to the qubit at time $t_r(t_i,\pm k_0)$ given by (see Appendix \ref{app-trt})
\begin{align}
    t_r(t_i,s k_0)=\!t_i+\left(\delta_{s,-1}+s\,\frac{k_0}{\pi}\right)T_B\,\,\,\,\,\,{\rm for}\,\,s=\pm1\,\,.\label{trpm}
\end{align}
Taking $t_r(t_i,\pm k_0)$ as the initial time of the next generated pair of wavepackets (one with momentum $k_0$ one with $-k_0$) the return times of these can be calculated again through \eq\eqref{trpm}. Iterating this procedure (starting from $t_i=0$) yields the series of all return times as a function of $\omega_0$ (see Appendix \ref{app-trt}).

The above discussion holds for $F$ weak enough and $\omega_0$ not too close to a band edge. Evidence of this is provided by \figs\ref{fig:sim}(d1)-(d3), where we reconsider the case $\omega_0=0$ as in panels (b1)-(b3) but now for a stronger force, and by \figs\ref{fig:sim}(e1)-(e3) where $\omega_0$ is parked in the vicinity of the upper band edge. In the first case, the trajectories of emitted wavepackets become less and less defined and, correspondingly, the qubit revivals increasingly wider and structured. In the latter case, zero plateaux in the qubit population not even occur, which can be understood from the fact that  wavepackets emitted to the left take a negligible time to return to the qubit. The distance traveled by such wavepackets is also vanishing, explaining why -- remarkably -- emission is almost fully chiral in this case.
\\
\\
A comprehensive analytical description of the complex emission dynamics in the weak-force regime is demanding and goes beyond the scope of the present work. In Section \ref{sec-dde}, we will however demonstrate that in the regime $\omega_{0}=0$ (qubit tuned at the band center) and $F$ weak enough the qubit decay is governed by a delay differential equation.

   \section{Theoretical analysis }\label{sec-dde}
   
   In line with the standard approach to treat light-matter interactions in photonic lattices
   in quantum optics and open quantum systems \cite{Lambropoulos2000a, breuer2002theory},
   it is natural to represent the dynamics in terms of the bare field's spectrum and eigenstates, which were reviewed in Section \ref{sec-eig-HB}.
   Resorting again to the conservation of the total number of excitations, the joint evolved state at any time $t$ during the emission process can also be written as
   \begin{equation}
   	\ket{\Psi(t)}={\alpha_e}(t) \ket{e}+\sum_n \alpha_n(t) \ket{\varphi_n}\label{psit}\,,
   \end{equation}
   where  $\alpha_n(t)$ is now the probability amplitude to find the system at time $t$ in state $\ket{\varphi_n}$ [recall \eq\eqref{phin}].  \eq\eqref{psit} differs from \eqs\eqref{psit1} and \eqref{psit2} in that it is expressed in terms of the field's eigenstates $\{\ket{\varphi_n}\}$.

   Replacing \eqref{psit} into the Schr\"odinger equation $i \frac{d}{dt}\!\ket{\Psi(t)}=H \ket{\Psi(t)}$ with $H$ given by \eq\eqref{Htot} gives rise to the system of differential equations (in a frame rotating at the qubit frequency $\omega_0$)
    \begin{align}
   	\dot{\alpha}_e=-i\sum_n g_n \,\alpha_n\,,\,\,\,\dot\alpha_n=-i (\omega_n-\omega_0) \alpha_n-i g_n {\alpha_e}\,\label{diff}
   \end{align}
   with $\omega_{n}$ the Wannier-Stark ladder spectrum given by \eq\eqref{ladder} [see also \fig\ref{fig:setup}(c)] and where we defined the coupling-mode function 
    \begin{equation}
     g_n=g\langle n_0| \varphi_n\rangle=g\,J_{n_0-n}(\xi)\,.\label{gn}
    \end{equation}

   \subsection{Regimes of light-matter interaction}\label{sec-regimes}
   
   Similarly to a multi-mode perfect cavity (which also features a ladder spectrum of normal frequencies) \cite{RotterVolterra14,SundaresanPRX15}, the number of field eigenstates $\ket{\varphi_n}$ which must be accounted for to describe the qubit emission depends on how large is the field energy {\it spacing} $\Delta \omega_n=F$ compared to the {\it characteristic strength} of the coupling mode function $g_n$. Recalling that each field eigenstate is localized in a region of length $\sim \xi$ [\cf\eq\eqref{xi}], the latter can be estimated as
   \begin{equation}
   	\bar g \sim  \frac{g}{\sqrt{\xi}}\,.\label{barg}
   \end{equation}
   Accordingly, 
   \begin{equation}
   	\frac{\bar{g}}{\Delta \omega_n}\sim \sqrt{\Gamma T_B}\,,
   \end{equation}
   where we recall that $\Gamma$ and $T_B$ are respectively the qubit decay rate for $F=0$ [\cf\eq\eqref{Gamma}] and the Bloch oscillations period [\cf\eq\eqref{TB}]. 
   
   Therefore, the key parameter is the dimensionless quantity $\Gamma T_B $, namely the ratio between the period of Bloch oscillations and the qubit decay time (for $F=0$). Accordingly, the strong (weak) force regime can be thus identified by the condition $\Gamma T_B \ll1 $ ($\Gamma T_B \gg1 $). The physical interpretation is that for $\Gamma T_B \ll1 $ the characteristic time taken by light to travel across the band is negligible compared to the qubit decay time. Conversely, for $\Gamma T_B \gg1 $, the qubit has enough time to decay to the ground state before a Bloch period is complete.

    \subsection{Strong-force: vacuum Rabi oscillations}\label{sec-rabi-th}
    
    For $\Gamma T_B\ll 1$ we are in the strong force regime, which is the one involved in Section \ref{sec-rabi} (indeed the parameters set in \fig\ref{fig:rabi} yield $\Gamma T_B\sim 10^{-5}$). Accordingly, by tuning the qubit frequency such that $\omega_{0}\simeq \omega_{n_c}$ for one selected field normal frequency $\omega_{n_c}$, all field normal modes with $n\neq n_c$ can be neglected and one reduces to an effective Jaynes-Cummings model where the cavity mode is embodied by the field eigenstate $\ket{\varphi_{n_c}}$. Interestingly, recall that 
    -- in general -- this mode has asymmetric spatial shape [\cf\eq\eqref{fig:bessel}] and, even more remarkably, is not spatially centered at the qubit location being indeed localized around cavity $n_c$ with characteristic width $\xi$ [see \eq\eqref{xi} and \fig\ref{fig:bessel}].
    These features are reflected in the form of the effective qubit-mode coupling strength which is given by [\cf\eq\eqref{gn}] $g_{c}= g\,J_{n_0-n_c}(\xi)$.
    In order for such interaction to be significant, the qubit must be thus placed within a distance $\sim\! \xi$ from the mode center, producing the condition $|n_0-n_c|\lesssim \xi$ (recall Section \ref{sec-eig-HB}). Hence, $\xi$ plays the role of the cavity length with the last condition saying the qubit must be placed inside this effective cavity in order to appreciably interact with it.
    
    From standard cavity QED theory \cite{haroche_exploring_2013}, the effective Rabi frequency is given by $\Omega=\sqrt{\delta^2+4g_c^2}$ with $\delta$ the detuning from the cavity mode, which in the present system explicitly reads 
    \begin{equation}
    	\Omega=\sqrt{(\omega_{0}-\omega_{n_c})^2+4g^2\,J^2_{n_0-n_c}(\xi)} \,
    \end{equation}
    with $\omega_{n_c}=n_c F$.
    
    For $n_0=0$, this correctly matches the frequency of vacuum Rabi oscillations in \fig\ref{fig:rabi} in the cases $\omega_0=0$, $n_c=0$ and $\omega_0=\pm 3F$, $n_c=\pm 3$. It is now apparent that for $n_c>n_0$ ($n_c<n_0$) and $\omega_{0}\simeq \omega_{n_c}$ the effective cavity mode lies on the right (left) of the qubit, explaining why the chirality of vacuum Rabi oscillations can be controlled by tuning the qubit frequency.

     \subsection{Weak-force: delay differential equation }\label{sec-dde}
     
     Let us now discuss the more involved weak-force regime $\Gamma T_B\gg 1$, in which case the qubit will see a dense set of field normal modes as discussed in Section \ref{sec-regimes}. This demands a multi-mode description of the dynamics based on the differential system \eqref{diff}. 
     
     For the sake of argument and in line with \figs\ref{fig:sim} and \fig\ref{fig:disp}, henceforth we will set $n_0=0$.
     Solving for each $\alpha_n(t)$ as a function of $\alpha_e(t)$ in the second identity of \eqs\eqref{diff} and replacing in the first identity gives rise to a closed integro-differential equation for $\alpha_e(t)$ only that reads
     \begin{equation}
     	\frac{d\alpha_e}{dt}=-g^2 \int_0^t\!d\tau\, K(\tau) \,\alpha_e (t-\tau)\,\label{MK-el}\,.
     \end{equation}
     Using $J_{-n}(\xi)=(-1)^n J_n(\xi)$ the memory kernel is given by
     \begin{equation}
     \begin{split}
     	{\cal K}(\tau) &= \sum_{n=-\infty}^{\infty} \,\left[J_{n}(\xi) \right]^2\, e^{-i (F n-\omega_0) \tau}\,,\label{MK}     
        \\
        & = e^{i\omega_0 \tau} J_0 \left[\xi \sin \left(\pi\tfrac{\tau}{T_B}\right)\right].
     \end{split}
     \end{equation}
     where in the second identity we worked out the series as a Bessel function of the first kind of order zero yet with an oscillatory argument (see \eg \rref\cite{Hartmann_2004}).
     The memory kernel function thus has a periodic behaviour with a period given by $T_B$, witnessing the intrinsically non-Markovian nature of the qubit decay. 
     
     \eq\eqref{MK-el} fully describes the qubit decay dynamics, but it is hard to treat in general due to the complicated form of the memory kernel function. Thus in order to infer information about the physics of the problem, approximations are needed in order to arrange the equation in a more intuitive form. We show next that this task can be accomplished in the case that the qubit frequency lies at the band center, \ie for $\omega_{0}=0$.

To show this, notice that, out of all the field eigenstates $\ket{\varphi_n}$, the qubit will predominantly interact with those of frequency such that $\omega_n\simeq \omega_0$ (amost resonant eigenstates) corresponding to slowly-varying terms in the first identity of \eqref{MK}. Now, since $\omega_n=n F$ [\cf\eq\eqref{ladder}], it turns out that these eigenstates are also such that $n\simeq 0$, i.e., they are the spatially nearest to the qubit location. More in detail, recalling \eq\eqref{barg}, these are all states $\ket{\varphi_n}$ fulfilling $|n|\lesssim n_{\rm max}$ where $n_{\rm max}$ can be estimated as  $n_{\rm max}= {\bar{g}}/{\Delta \omega_n}$. Now, since $g\ll J$ (as we assume throughout this work), it turn out that $ {\bar{g}}/{\Delta \omega_n}\ll \sqrt{\xi}$ \footnote{Indeed, with the help of \eq\eqref{xi} we get $\frac{\bar{g}}{\Delta \omega_n}\sim \frac{g}{\sqrt{2J F} }\ll \frac{J}{\sqrt{2J F} }\sim \sqrt{\xi} $.}. 
Hence, we find that almost resonant eigenstates in this case satisfy the condition $|n|\ll 	\sqrt{\xi}$. This means that the qubit lies far enough from the tails of the wavefunction of each of such eigenstates, allowing to approximate each of them as having a sinusoidal shape based on \eq\eqref{sin}. Accordingly, in this regime the memory kernel \eqref{MK} can be effectively approximated as
\begin{equation}\label{K2}
 	    \begin{split}
     	{\cal K}(\tau)&\simeq \frac{2}{\pi \xi}\sum_{n=-\infty}^{\infty}  \,\sin^2 \left(n \frac{\pi}{2}+\xi+\frac{\pi}{4}\right)\, e^{-i {F} n \tau}\,
     \end{split}
\end{equation}       
By decomposing next the sine function into complex exponentials, this can be equivalently written (see Appendix \ref{app-dde}) as a sum of Dirac delta functions 
\begin{equation}\label{K3}
\begin{split}
		{\cal K}(\tau)=\frac{T_B}{\pi\xi}\,\Big[&\sum_{\ell=0}^\infty \delta(\tau-\ell T_B)\\\,\,\,&+\sin{(2\xi)}\sum_{\ell=0}^\infty \delta[\tau-(\ell+\frac{1}{2}T_B)] \Big].
\end{split}
\end{equation}  
Replacing this back into \eq\eqref{MK-el} one eventually ends up with the non-Markovian differential equation governing the qubit emission:
\begin{align}
	\frac{d\alpha_e}{dt}=&-\frac{\Gamma}{2} \alpha_e (t)-\Gamma \sum_{\ell=1}^\infty \alpha_e (t-\ell T_B) \Theta\left(t{-}\ell T_B\right)\nonumber\\
	&-\Gamma  \sin(2\xi) \sum_{\ell=0}^\infty \alpha_e [t-(\ell{+}\tfrac{1}{2}) T_B]\, \,\Theta [t-(\ell{+}\tfrac{1}{2}) T_B]\label{dde2}
\end{align}
with $\Theta(x)$ the Heaviside step function.
Mathematically, \eq\eqref{dde2} is an instance of delay differential equation \cite{driver2012ordinary} whose corresponding memory kernel function peaks at multiple integers of $T_B/2$ which coincide with the return times $t_r$ of emitted wavepackets during the dynamics (see Section \ref{sec-NM}). The dynamics governed by \eq\eqref{dde2} can be worked out by solving exactly the generally inhomogeneous differential equation in each time interval $t\in [\ell T_B/2, (\ell+1) T_B/2[$ and using the resulting solution to define the inhomogeneous term to solve the equation in the next time interval \cite{driver2012ordinary}. We checked that the full solution of \eq\eqref{dde2} resulting from this iterative procedure accurately reproduces the one obtained from numerical simulations provided that $F$ is small enough [as is \eg the case in \fig \ref{fig:sim}(b1)].  
For times shorter than $T_B/2$, only the first term in the right-hand side of \eq\eqref{dde2} is non-zero: in this initial time interval, the qubit is effectively unaware of the presence of the electric field and undergoes the standard exponential decay according to $\alpha_e(t)=e^{-\frac{\Gamma}{2}t}$. At time $t=T_B/2$, the first revival generally occurs [see second line of \eq\eqref{dde2}] corresponding to the first return time of the two previously emitted wavepackets. Interestingly, notice that in \eq\eqref{dde2} terms corresponding to odd multiple integers of $T_B/2$ (second line of the equation) can identically vanish whenever $\xi=\nu \pi/2$ corresponding to the force values $F=4J/(q \pi)$ with $q$ an integer number, which is due to destructive interference of two emitted wavepackets as they return to the qubit. In these cases, qubit revivals at odd multiple integers of $T_B/2$ do not occur, which we numerically checked.

\subsection{Connection with a multi-mode cavity}    

\eq\eqref{dde2} is formally analogous to the delay differential equation \cite{milonni_PhysRevA.35.5081} (see also \rref\cite{milonni1983exponential} corresponding to the case where $\sin (2\xi){=}0$] governing the decay of an atom placed in the middle of a perfect cavity in the regime where, differently from the usual Jaynes-Cummings model, one has to take into account the multi-mode nature of the cavity and, consistently, the time delay $t_d$ taken by a photon to travel between the two cavity mirrors (with $t_d$ scaling as $L/v$ where $L$ is the cavity length and $v$ the photon speed). The role of $t_d$ and $L$ in our system are played by $T_B$ and $\xi$, respectively. 
Such mapping to a multi-mode cavity is a consequence of the ladder energy spectrum of normal modes combined with the sinusoidal approximation of each mode for $\omega_{0}=0$ [recall \eq\eqref{K2} and related discussion]. Interestingly, in a cavity the non-Markovian decay with revivals originates from the spatial confinement due to the actual cavity mirrors. Instead, in the present mirrorless setup it is caused by the combined effect of the synthetic force and band finiteness: the former makes the emitted photon change its momentum and energy in time while the latter constrains this motion through band edges. A more thorough comparison between the present system and an atom in a multi-mode cavity is presented in Appendix \ref{app-gdde}.

\section{Experimental implementations}\label{sec-exp}
 The predicted phenomena can be potentially implemented in any state-of-the-art waveguide-QED platform enabling the fabrication of a 1D photonic lattice coupled to quantum emitters \cite{Chang_RevModPhys.90.031002}. Two particularly promising implementations are superconducting arrays of resonators coupled to  transmon qubits \cite{gasparinetti_PhysRevX.12.031036, Painter_2018, Painter_PhysRevX.11.011015} and matter-wave emulators employing cold atoms in optical lattices \cite{Schneble_nature2018_matterwaves,daley_PhysRevLett.101.170504, GonzalezTudela2018nonmarkovianquantum, Schneble_PhysRevResearch.2.043307, devega_PhysRevLett.101.260404, Navarrete-Benlloch_2011}.
 In these setups, high-quality factors and very limited dissipation ensure that \eq \eqref{Htot} captures the whole qubit-field dynamics with very good approximation.

 While the strong-force regime entailing vacuum Rabi oscillations is straightforward to achieve, the weak-force regime, where non-Markovian decay with revivals occur, demands the gradient of cavity frequencies (\ie, the force) to be small but yet much larger than the unavoidable disorder.
 In state-of-the-art superconducting coupled-resonator arrays \cite{Painter_2018, Simon_2019_nature, Painter_PhysRevX.11.011015}, for instance, typical ranges of the hopping rate  and coupling strength are respectively $J/(2\pi)\sim 100{-}300\,$MHz and $g/(2\pi)\sim 1{-}100\,$MHz, but due to fabrication imperfections the frequency of each resonator is subject to disorder with a typical uncertainty  $\sigma_{\omega}/(2\pi)\sim 1\,$MHz which thus imposes the requirement $F>\sigma_{\omega}$.
 As an instance, a device with $g/(2\pi)\sim 50\,$MHz, $J/(2\pi)\sim 300\,$MHz and $F/(2\pi)\sim 5\,$MHz would match this condition and appears within reach or at least not far-fetched.

\section{Conclusions}\label{sec-conc}

In summary, in this work we investigated the decay of a qubit into a photonic lattice where emitted photons are accelerated by a synthetic force. Despite the qubit being weakly coupled to the array, the force drives the emitted photon along the band until hitting one of the two band edges. Somewhat similarly to a mirror, despite the absence of actual mirrors in the system, this causes an inversion of motion enforcing the emitted photon to return to the qubit where, depending on interference conditions, it can be generally reabsorbed. The non-Markovian dynamics seeded by this novel feedback mechanism is rich and generally involved. Here, we provided a first characterization by identifying in particular two regimes corresponding to strong and weak values of the synthetic force. In the former case, we showed that vacuum Rabi oscillations can occur with the qubit periodically exchanging energy with a field eigenmode spread over many cavities of the array. These can lie either to the right or left of the qubit, meaning that chiral reversible emission occurs whose chirality can be controlled by tuning the qubit frequency. For weak force, the qubit instead  exhibits a complex non-Markovian decay with revivals where each emitted wavepacket undergoes a round trip before returning to the qubit. When the qubit is tuned to the band center, we proved that such dynamics is described by a delay differential equation formally analogous to the one describing the decay of an atom into a multi-mode standard cavity.

The present study links a very distinctive and longstanding effect of having a band structure (Bloch oscillations) with non-Markovian dynamics in a 1D wavevuide. It is remarkable that the usual in-band exponential decay occurring for weak coupling and no applied force does not show up any direct sign of the presence of photonic band edges. In contrast,  turning on even a weak force makes the qubit aware that the field lives in a finite-size band, which deeply affects its dynamics. This occurs because the emitted photon is accelerated by the synthetic force; as such, its energy and momentum are no longer constant during the evolution. Such kind of dynamics is to our knowledge unprecedented in waveguide QED and thus has the potential to introduce a new  paradigm of atom-photon interactions, including implementations of quantum information processing tasks.
 
Somewhat in line with several papers investigating non-Markovian decay with time delays in standard waveguides (see \eg \rrefs \cite{DornerPRA02,TufarelliPRA13,GrimsmoPRL15,pichlerPhotonic2016}), in this work we focused on emission properties of a single qubit, which was shown to be an already quite involved dynamics. The extension to many qubits is an interesting outlook, for instance in order to devise novel entanglement generation schemes, but demands introducing extra degrees of freedom and an additional characteristic length (the distance between the qubits) into the problem; as such, it goes beyond the scope of the present paper and will be the subject of future investigations.

\acknowledgments

MP, GLS and FC acknowledge financial support from European Union-Next Generation EU through projects: Eurostart 2022 ``Topological qubit-photon interactions for quantum technologies"; PRIN 2022–PNRR No. P202253RLY ``Harnessing topological phases for quantum technologies"; THENCE–Partenariato Esteso NQSTI–PE00000023–Spoke 2 ``Taming and harnessing decoherence in complex networks".
DDB acknowledges funding from the European Union - NextGeneration EU, "Integrated infrastructure initiative in Photonic and Quantum Sciences" - I-PHOQS [IR0000016, ID D2B8D520, CUP B53C22001750006].

\appendix

\section{Return times}\label{app-trt}

Consider a wavepacket emitted from the qubit at time $t_i$ whose average position evolves as in \eq\eqref{xt}. The return time $t_r$ is the earliest non-zero time such that $x(t_r)=x_i$. This fulfils the condition
\begin{equation}
k_i- 2\pi\frac{t_r-t_i}{T_B}=-k_i+ 2\nu \pi\,,
\end{equation}
with $\nu$ an integer number,
which gives
\begin{equation}
t_r-t_i=\frac{k_i}{\pi}\,T_B+\nu T_B\,.
\end{equation}
Taking into account the constraint $t_r-t_i\ge0$, we get 
\begin{align}
	t_r(t_i,k_i)&=t_i+\left(1+\frac{k_i}{\pi}\right)T_B\,\,\,\,\,{\rm for}\,\,-\pi< k_i< 0\,,\\
	t_r(t_i, k_i)&=t_i+\frac{k_i}{\pi}\,T_B\,\,\,\,\,{\rm for}\,\,\,\,0\le k_i\le  \pi\,,
\end{align}
where on the left-hand side we have highlighted the dependence on $t_i$ and $k_i$.
Note that at these times not only the wavepacket average position is back to the initial value $x_i$, but even its average energy takes again the initial value $\omega_i=-2J \cos k_i$ [since the oscillations in \eqs\eqref{xt} and \eqref{wt} are in-phase].

The whole set of return times during the emission dynamics is obtained iteratively as follows. The return times of the first two emitted wavepackest are given by $t_r(0,\pm k_0)$. Either of these is the initial time at which a new pair of  wavepackets of momenta $\pm k_0$ are emitted (thus four overall). By iterating this procedure, the whole set of return times can be predicted. In the special case $k_0=\pm \pi/2$, it is easy to see that the set of return times are simply $\nu T_B/2$ with $\nu=1,2,..$. 

Notice that the above computation scheme of return times implicitly assumes that each wavepacket is emitted and absorbed instantaneously. This picture breaks down at long times for $\omega_0\neq 0$, as \eg in \fig\ref{fig:sim}(c1), (c2) and (c3), since the return times of emitted wavepackets get more and more dense.

\section{Derivation of the delay differential equation}\label{app-dde}

Upon decomposition of the sine function into complex exponentials, \eq\eqref{K2} can be expressed as
\begin{align}
	F(\tau)=\frac{1}{\pi \xi} \sum_{n=-\infty}^{\infty} e^{-i F n \tau}&+\frac{i}{2\pi \xi} e^{-i 2\xi}\sum_{n=-\infty}^{\infty} e^{i (\pi-F \tau)n}\nonumber\\
	&-\frac{i}{2\pi \xi}e^{i 2\xi} \sum_{n=-\infty}^{\infty} e^{-i (\pi+F \tau)n}\,\label{Ftau}
\end{align}
With the help of the Poisson summation formula
\begin{equation}\label{Poisson}
	\sum_{n=-\infty}^\infty e^{i n x}=\sum_{n=-\infty}^\infty \delta (x/(2\pi)-n)\,
\end{equation}
the three series in \eqref{Ftau} can be conveniently expressed in terms of Dirac delta functions as
\begin{align}
	\sum_{n=-\infty}^\infty e^{-i {F} n \tau}&=\frac{2\pi}{{F} }\sum_{n=-\infty}^\infty \delta\left(\tau+n\frac{ 2\pi}{{F}}\right)\,,\\
	\sum_{n=-\infty}^\infty e^{i (\pi-{F} \tau)n}&=\frac{2\pi}{{F} }\sum_{n=-\infty}^\infty \delta\left[\tau+(n-1/2) \frac{2\pi}{{F}}\right]\,,\\
	\sum_{n=-\infty}^\infty e^{-i (\pi+{F} \tau)n}&=\frac{2\pi}{{F} }\sum_{n=-\infty}^\infty \delta\left[\tau+(n+1/2)\frac{ 2\pi}{{F}}\right]\,.
\end{align}
Taking into account that in our case $\tau$ is anyway positive, these can be equivalently arranged as (note in particular the lower bounds of summation)
\begin{align}
	\sum_{n=-\infty}^\infty e^{-i {F} n \tau}&=\frac{2\pi}{{F} } \,\left[\delta(\tau)+\sum_{n=1}^\infty \delta\left(\tau-n \frac{2\pi}{{F} }\right)\right]\,,\!\!\!\!\!\label{sum1}\\
	\sum_{n=-\infty}^\infty e^{i (\pi-{F} \tau)n}&=\frac{2\pi}{{F} }\sum_{n=0}^\infty \delta\left[\tau+(n+1/2)\frac{2\pi}{{F} }\right]\,,\\
	\sum_{n=-\infty}^\infty e^{-i (\pi+{F} \tau)n}&=\frac{2\pi}{{F} }\sum_{n=1}^\infty \delta\left[\tau+(n-1/2) \frac{2\pi}{{F} }\right]\,.
\end{align}
Plugging these back into \eq\eqref{Ftau} with helpf of \eqs\eqref{xi}, \eqref{TB} and \eqref{Gamma} leads to \eq\eqref{K3}.

Finally, once $K(\tau)$ is replaced with \eqref{K3} in \eq\eqref{MK-el} and recalling that $\int_0^t \delta (\tau-\tau^*) f(\tau)=f(\tau^*)\Theta (t-\tau^*)$ and $\Theta(0)=1/2$, one ends up with the delay differential equation \eqref{dde2}.

\section{Delay differential equation for generalized coupling-mode function}\label{app-gdde}

Rigorously speaking, despite sharing a ladder-type photonic spectrum, our system differs from an atom coupled to a multi-mode cavity in the functional form of the coupling-mode function $g_n$. Notwithstanding, the atomic excitation amplitude is governed by the same kind of delay differential equation. This indeed holds for a large class of photonic baths with   ladder spectrum, as shown by the following general argument.

Consider a photonic bath with a ladder spectrum of normal frequencies $\omega_n=n\Delta$, where $n$ labels the modes and $\Delta$ is the energy spacing (in our case $\Delta=F$). A two-level emitter with frequency $\omega_0$ is coupled to the field under the rotating wave approximation so that \eqs\eqref{diff} still hold but with the coupling-mode function now taking the  general form
\begin{equation}\label{couplingmodefunction}
    g_n=g\, f(n) \sin\left(n\frac{\pi}{2}+\varphi\right)\,.
\end{equation}
In our system for $\omega_0=0$ and $F$ weak [\cf\eqs\eqref{gn} and \eqref{sin}] $f(n)$ is $n$-independent, whereas for a standard multi-mode cavity  $f(n)\propto \sqrt{n}$.

Through the mild requirement that function ${\cal{F}}(n)=f^2(n)$ be analytic, the integro-differential equation governing the evolution of the atomic amplitude $\alpha_e(t)$ in the weak-coupling regime  reads
\begin{equation}
\begin{split}
    \frac{d\alpha_e}{dt}=-g^2\int_{0}^{t} dt^\prime \sum_{n=-\infty}^{\infty} {\cal{F}}(n)\sin^2\left(\frac{n\pi}{2}+\varphi \right)\\\,\,\,\,\,\,\,\,\,\,\,\times \,e^{-i(\omega_n-\omega_0)(t-t^\prime)}\alpha_e(t^\prime)\,.
\end{split}.
\end{equation}
By decomposing the sine function into complex exponentials and upon a Taylor-expansion of ${\cal{F}}(n)$, each power of $n$ can be conveniently expressed as a time derivative of the complex exponential according to
\begin{equation}
\begin{split}
    \frac{d\alpha_e}{dt}=&-g^2\int_{0}^{t} dt^\prime e^{i\omega_0(t-t^\prime)} \sum_{m=0}^{\infty} \frac{{\cal{F}}^{(m)}(0)}{m!}\frac{1}{(-i\Delta)^m}\\
   &\,\,\,\frac{d^m}{dt^m} \sum_{n=-\infty}^{\infty}e^{-in\Delta (t-t^\prime)}\sin^2{\left(\frac{n\pi}{2}+\varphi \right) \alpha_e(t^\prime)}\,.\label{diff-gen}
\end{split}
\end{equation} 
By applying the Poisson summation formula \eqref{Poisson} along with the  property of Dirac delta functions
\begin{equation}
    \frac{d^m}{dt^m}\delta(t-t^*)=(-1)^m\delta(t-t^*)\frac{d^m}{dt^m}\,,
\end{equation}
\eq\eqref{diff-gen} can be turned into the delay differential equation 

\begin{align}
	\frac{d\alpha_e}{dt}&=-\frac{g^2 T}{2}f^2\left(\frac{\omega_0}{\Delta}\right)\Big[ \frac{\alpha_e(t)}{2}{-} \cos(2\varphi)\nonumber\\
	&\qquad\times \sum_{n=0}^{\infty}e^{i\omega_0 T(n{+}\tfrac{1}{2})}\alpha_{e}[t{-}T(n{+}\tfrac{1}{2})]\Theta[t-T(n{+}\tfrac{1}{2})] \Big]\label{dde3}
\end{align}
where we set $T=2\pi/\Delta$.

In our case, $f(n)=\sqrt{2/(\pi \xi)}$, $T=T_B=2\pi/F$, $\varphi=\xi+\frac{\pi}{4}$ and $\omega_0=0$ [cf. Eqs. \eqref{couplingmodefunction}, \eqref{sin},\eqref{TB} and \eqref{gn}], in which case \eq\eqref{dde3} reduces to \eq\eqref{dde2}.
 
\bibliography{PSDBC_v1} 
\end{document}